\documentclass[fleqn,11pt]{article}
\usepackage{microtype}
\usepackage{setspace}
\usepackage{hyperref} 
\usepackage{arxiv}
\usepackage{subcaption}
\usepackage[utf8]{inputenc} 
\usepackage[T1]{fontenc}    
\usepackage{url}            
\usepackage{booktabs}       
\usepackage{amsfonts}       
\usepackage{nicefrac}       
\usepackage{microtype}      
\usepackage{lipsum}         
\usepackage{graphicx}
\usepackage{cite}
\usepackage{doi}
\usepackage{amsmath,amssymb,amsfonts}
\usepackage{algorithmic}
\usepackage{graphicx}
\usepackage{textcomp}
\usepackage{xcolor}
\def\BibTeX{{\rm B\kern-.05em{\sc i\kern-.025em b}\kern-.08emT\kern-.1667em\lower.7ex\hbox{E}\kern-.125emX}}

\usepackage{booktabs} 
\usepackage{graphicx}
\usepackage{epstopdf}
\usepackage{url}
\usepackage{algorithmic}
\usepackage{multirow}
\usepackage{caption}
\usepackage{amssymb}
\usepackage{amsfonts}
\usepackage{float}
\usepackage{bm}
\usepackage{dsfont}
\usepackage[T1]{fontenc}
\usepackage[left,modulo]{lineno}
\usepackage[ruled, vlined, linesnumbered, longend]{algorithm2e}
\usepackage{psfrag}
\usepackage{fancyhdr}
\usepackage{eucal}
\usepackage[english]{babel}
\usepackage{ifpdf}
\usepackage{cleveref} 
\usepackage{textcomp}
\usepackage{tikz}
\usepackage{tikz-inet}
\usetikzlibrary{calc,shapes.geometric}
\usetikzlibrary{shapes,snakes,arrows,automata} 
\usetikzlibrary{patterns}
\usepackage{multirow}
\usepackage{multicol}
\usepackage{authblk}
\begin{document}
%

%

\title{Physics-Informed Mixture Models and Surrogate Models for Precision Additive Manufacturing}
%
%
%
\newbox{\orcid}\sbox{\orcid}
{\includegraphics[scale=0.06]{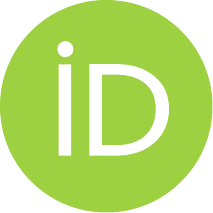}} 
%
\author{
\parbox{\textwidth}{
  \href{https://orcid.org/0000-0002-9172-0155}{\includegraphics[scale=0.08]{orcid.pdf}}\hspace{1mm}Sebastian Basterrech, Shuo Shan, 
    \href{https://orcid.org/0000-0002-2254-7846}{ \includegraphics[scale=0.08]{orcid.pdf}}Debabrata Adhikari, 
   \href{https://orcid.org/0000-0002-1142-0989}{ \includegraphics[scale=0.08]{orcid.pdf}}Sankhya Mohanty\\
  {Department of Civil and Mechanical Engineering\\
  Technical University of Denmark,  Kongens Lyngby, Denmark}\\
  Emails: \texttt{sebbas@dtu.dk}, \texttt{sshan@dtu.dk}, \texttt{debaa@dtu.dk},
  \texttt{samoh@dtu.dk}
}}
\date{}
\renewcommand{\undertitle}{To appear in the proceedings of AI in Science Summit (AIS 2025), Copenhagen, Denmark}
\renewcommand{\shorttitle}{Physics-Informed Mixture Models and Surrogate Models for Additive Manufacturing}


\maketitle

\begin{abstract}
In this study, we leverage a mixture model learning approach to identify defects in laser-based Additive Manufacturing (AM) processes. 
By incorporating physics based principles, we also ensure that the model is sensitive to meaningful physical parameter variations.
%
%
The empirical evaluation was conducted by analyzing real-world data from two AM processes: Directed Energy Deposition  and Laser Powder Bed Fusion. In addition, we also studied the performance of the developed framework over public datasets with different alloy type and experimental parameter information.
The results show the potential of physics-guided mixture models to examine the underlying physical behavior of an AM system.
\end{abstract}

\section{Introduction}
Over the last decades, precision Additive Manufacturing (AM) has enabled significant advances in the adaptation of new materials for structural applications, and innovative and personalized designs that are infeasible to manufacture using traditional methods.
%
%
The AM processes are versatile as they have broad applicability with respect to the range of materials, such as metals, polymers and ceramics, that can be processed using the same machine with minimal modifications.
In addition, AM technologies bring the opportunity to customize the products according to specific demands by integrating a high level of digitalization into the production process. There remains however the potential to make the whole process more sustainable and cost-efficient, facilitated through simulations and digital twin implementations.
%
%
%
Recent methodological approaches aim to characterize a mapping between process-level information, the emergent material microstructure and the consequent material properties (e.g. mechanical, thermal, electrical and chemical properties).  
%
%
By specifying these functional relationship in a robust manner, we can develop an efficient process for creating products with controlled local  material properties.
%
%
%
%
%
%
%
%
%

Physics-informed mixture models form an emerging area with applications in industrial design processes. 
These techniques integrate domain knowledge into the well-known stochastic modeling workflows e.g. Gaussian Mixture Models (GMMs)~\cite{Xu1996} or Mixture of Experts~\cite{Kang2025pig}.
%
%
Physics-informed models overcome some limitations of classical machine learning (ML), in that they often require less data than large models, they generalize better, and they have enhanced physical  plausibility, interpretability, and parsimony~\cite{Meng2025}.
Physics-informed ML not only enhances the interpretability of the ML models, they also offer valuable insights into the physical systems.
In recent years, aided by the development of automatic differentiation techniques, physics-informed ML methods have been successfully used for solving multiphysics problems involving coupled partial differential equations~\cite{Du2025}.
%
%
%
%
%
%
%

In this work, we investigate the potential of physics-informed GMMs for detecting defects in two distinct types of AM processes: Laser Powder Based Fusion (L-PBF) and Directed Energy Deposition (DED).
%
%
The remainder of this short article presents the background concepts, describes the methodology, and discusses the results along with future research challenges

\section{Preliminaries}

\newcommand{\Prob}{\mathsf{P}}
\newcommand{\param}{\mathbf{x}}
\newcommand{\surface}{\mathbf{s}}
\newcommand{\paramDomain}{\mathcal{P}}
\newcommand{\surfaceDomain}{\mathcal{S}}

\subsection{Problem specification}
Let $\paramDomain$ denote the space of process-level parameters, for example powder size and shape, recoater velocity, scanning strategies and thermal parameters,  and let $\surfaceDomain$ be the space of  final properties, such as surface texture, defects (e.g. balling), and other bulk properties (e.g. mechanical strength).
We aim  to establish a surrogate relationship $\psi_{\bm{\theta}}(\cdot)$ parametrized by $\bm{\theta}$ such that for any given value of process parameters $\param \in \paramDomain$, the corresponding material properties $\surface = \psi_{\bm{\theta}}(\param)$, with $\surface\in\mathcal{S}$, can be  predicted and controlled.
In practice, this problem is formulated as an optimization task where, given a collection of $N$ input-output pairs $\{(\param_i,\surface_i)\}^N_{i=1}$, the goal is to find the parameter vector $\theta^*$ that minimizes the average discrepancy between the predictions of the surrogate function $\psi_{\bm{\theta^*}}(\param)$ and the true values $\surface$.\\
\subsection{Mixture models: the Gaussian case}
%
%
It assumes that the data is generated from a mixture of a fixed number of Gaussian distributions. Let $M$ denote the number of Gaussian components, the model is a mixture density of the form:~\cite{Hastie2001} 
\begin{equation}
\label{GMMform}
\Phi(\mathbf{x},\bm{\theta}) = \sum_{m=1}^M \alpha_m \, \phi(\mathbf{x};{\mu}_m, {\bm{\Sigma}}_m),
\end{equation}
where $\bm{\theta}$ collects the parameters $\{\bm{\alpha}, \bm{\mu},\bm{\Sigma}\}$. 
The elements $\alpha_m$ of the vector $\bm{\alpha}$ denote the mixing proportions, which sum up to~$1$. 
Each component has a mean $\mu_m$ and covariance matrix ${\bm{\Sigma}}_m$ forming the vectors $\bm{\mu}$ and $\bm{\Sigma}$, respectively.
%
%
For any given $\param$, the probability margin given by the expression~(\ref{GMMform}) can be used to estimate the likelihood of $\param$ under the model, i.e., to assess how well the GMM fits the data.
The parameters are usually adjusted using the Expectation-Maximization (EM) algorithm~\cite{Hastie2001}. 
%
%
GMMs are used in a wide range of problems as  density estimation to clustering problems.
When GMMs are used for solving classification tasks, a common approach is to apply generative classification. This involves training a GMM for each class and applying Bayes rule to compare the posterior probabilities. 
The decision rule consists of returning the class that maximizes the posterior probabilities.
Considering $K$ classes, the model $\Phi_k(\cdot)$ trained with the data with label $C_k$ has parameters $\bm{\theta_k}$, 
%
the  posterior probability of a class $C_k$ given an input $\param$ is defined as~\cite{Hastie2001}:
\begin{equation}
\label{Bayes}
\Prob(C_k \mid \param) = \frac{\Phi_k(\param, \bm{\theta}_k) \, \Prob(C_k)}{\sum_{j=1}^K \Phi_j(\param, \bm{\theta}_j) \, \Prob(C_j)}.
\end{equation}
This can be denoted as a vector of soft predictions for each class  $\hat{\bm{p}}(\param)=(\hat{p}_1(\param),\ldots,\hat{p}_K(\param))$; and the decision rule is to return the class that satisfies the $\arg\max_{k} p_k(\bm{x})$~\cite{Hastie2001}.
Another useful property of the GMMs is that generally produce smoother decision boundaries compared to other clustering tools such as K-means and Learning Vector Quantization~\cite{Hastie2001}.
%
%
%
%

\section{Material and methods}
%
\subsection{Embedding physics via surrogate features}
%
%
Once the task is well-specified, there are several ways of integrating physics into the learning framework, including feature selection, model architecture design, and the formulation of the loss function.
A common practice is to combine the raw data and surrogate models to map process parameters to thermal history and different mechanistic variables~\cite{Du2025,Saunders2023}.
In the current study, we introduce a normalized ``energy'' assessment defined as a non-linear combination of the laser power, scanning speed, and the specific heat capacity of the alloy. This function is inspired by the normalized enthalpy that is a ratio between the laser power and a combination of factors that govern heat absorption and diffusion in the material~\cite{Ghasemi-Tabasi2020}.
The normalized ``energy'' is defined as $P/(C_p V)$, where $P$ denotes the power, $V$ the scanning speed, and $C_p$ the specific heat capacity.
%
%
%
%
Strictly speaking, this is not a measure of energy; however, it serves as a proxy for it. 
It gives to the learning model an input feature with an approximate measure of how the energy is absorbed and distributed within the material. 
The selected input data related to the power exhibit a unimodal density, and the density related to the scan speed shows a bimodal shape in the case study with DED process.
Therefore, the normalized ``energy'' helps to transform the  data distribution into a unimodal form, which reduces complexity and helps the GMMs to focus in the meaningful structure.\\

%
\begin{figure}[H]
    \centering
    \begin{subfigure}[b]{0.48\textwidth}
        \centering
        \fbox{\includegraphics[width=7.6cm, height=4.75cm]{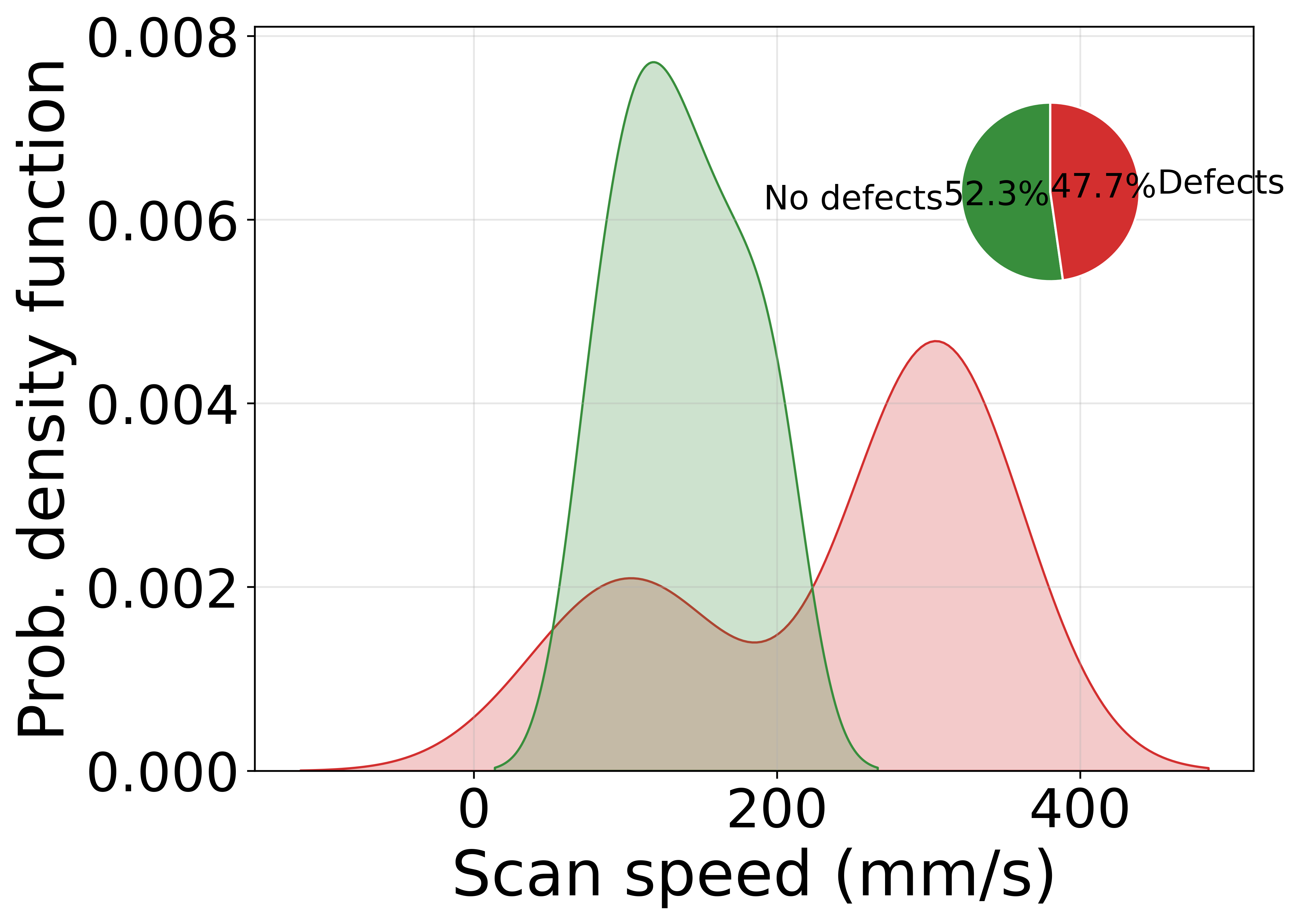}}
        \caption{Scan speed distribution in L-PBF data.}
        \label{fig:sub1}
    \end{subfigure}
    \hfill
    \begin{subfigure}[b]{0.48\textwidth}
        \centering
        \fbox{\includegraphics[width=7.6cm, height=4.75cm]{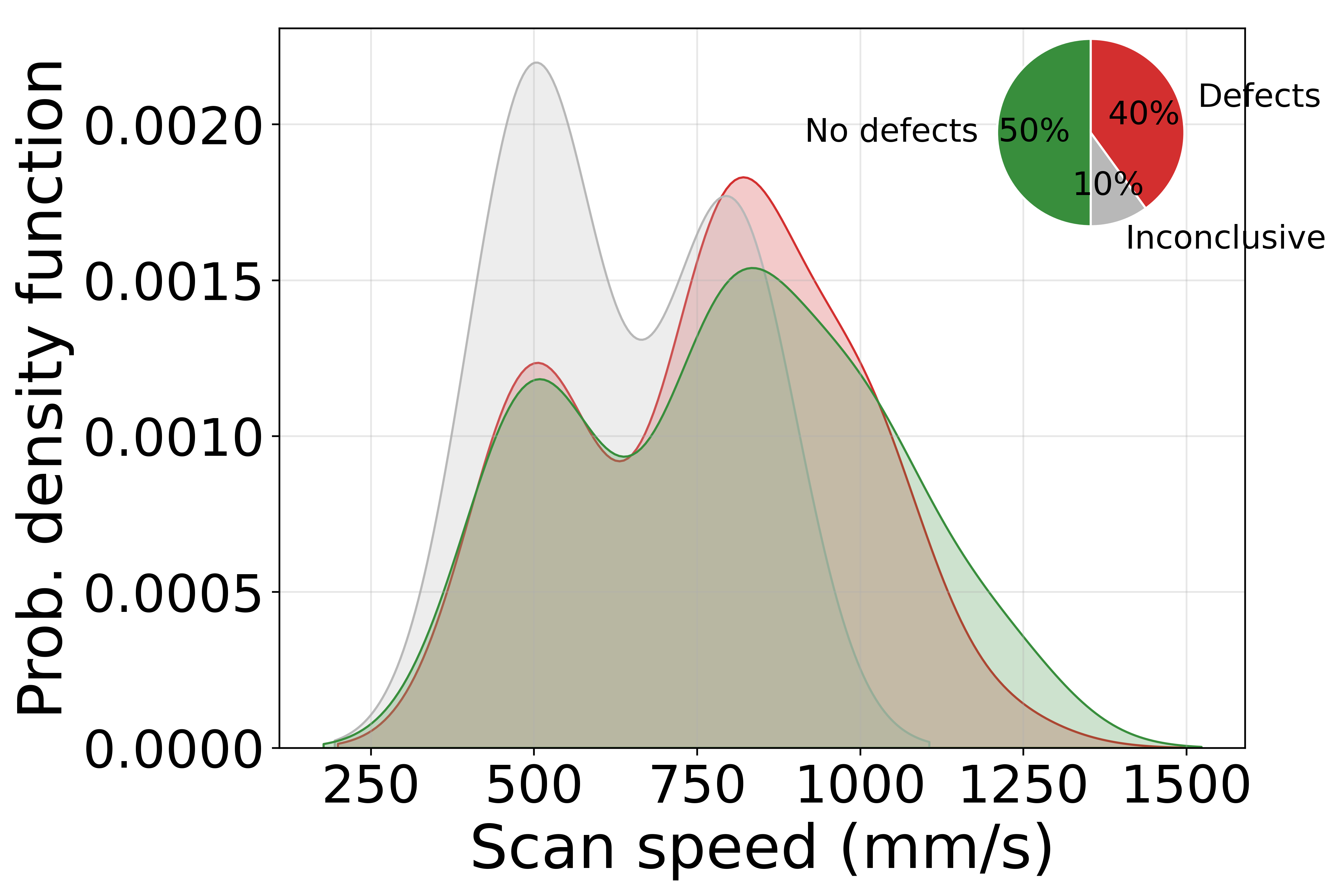}}
        \caption{Scan speed distribution in DED data.}
        \label{fig:sub2}
    \end{subfigure}
    
    \vspace{0.5cm}
    
    \begin{subfigure}[b]{0.48\textwidth}
        \centering
        \fbox{\includegraphics[width=7.6cm, height=4.75cm]{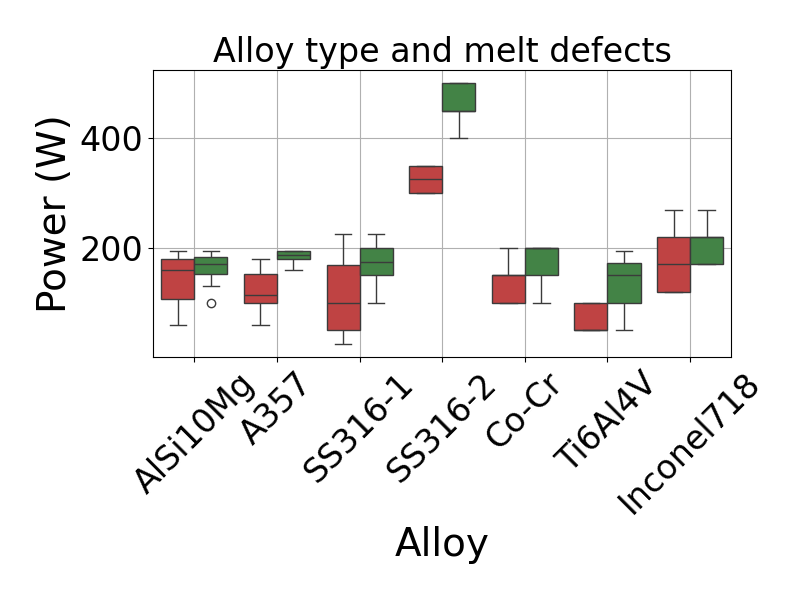}}
        \caption{Box plots showing the distribution of laser power for different alloy types.}
        \label{fig:sub3}
    \end{subfigure}
    \hfill
    \begin{subfigure}[b]{0.48\textwidth}
        \centering
        \fbox{\includegraphics[width=7.6cm, height=4.75cm]{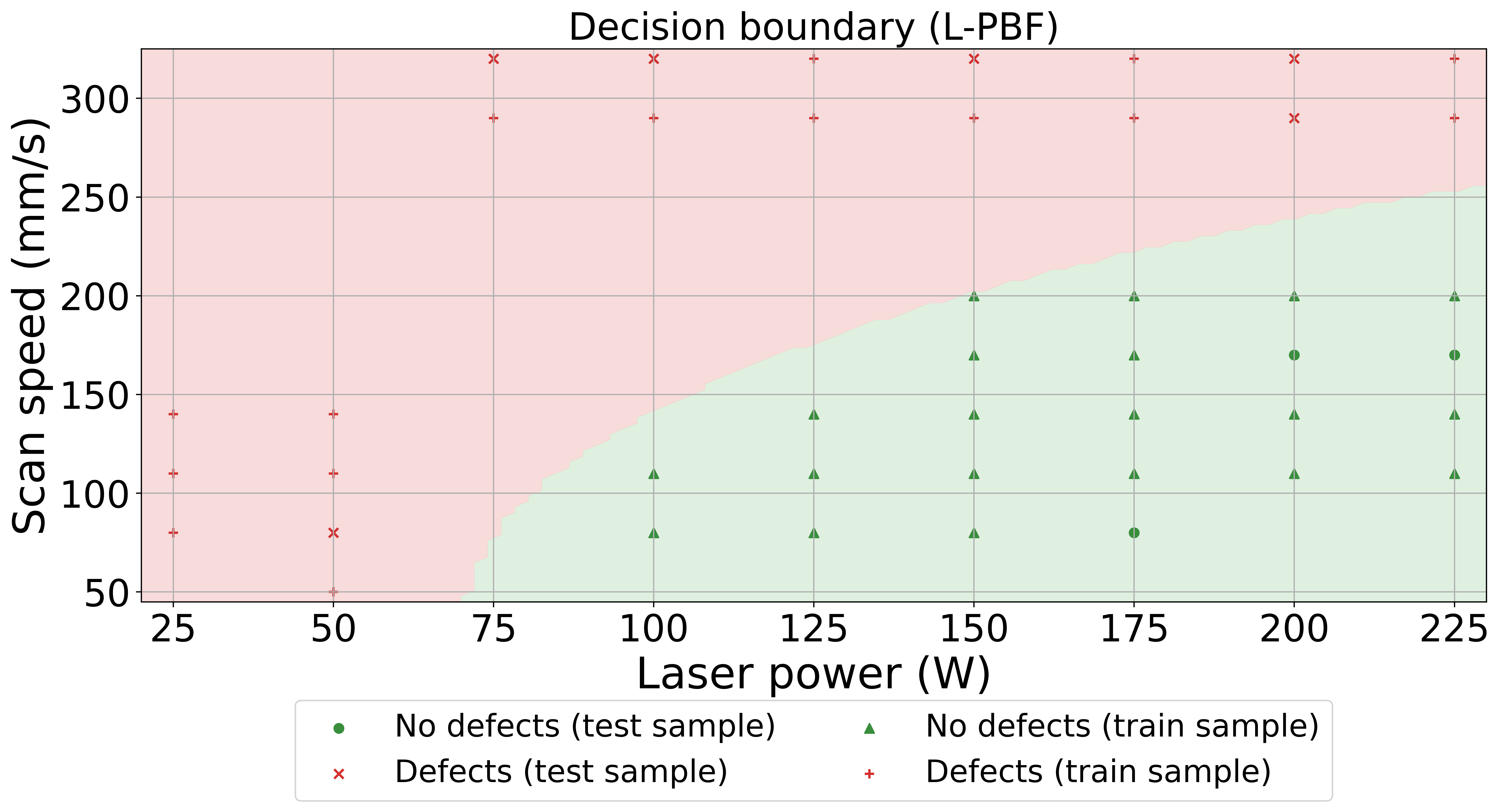}}
        \caption{Visualization of the decision boundary produced by the GMM model trained on L-PBF data. }
        \label{fig:sub4}
    \end{subfigure}
    
    \caption{Example of the data distribution and classification results.}
    \label{fig:results}
\end{figure}

\subsection{Study cases}
We evaluated the capability of physics-informed GMMs as defect detectors in two study cases: L-PBF and DED processes. The analyses were conducted with fabricated samples received from industrial production sites, with the stainless steel 316L (SS316-1) alloy type processes with L-PBF and the Inconel 625 processed by the DED process. 
%
%
%
The two processes differ primarily in how the molten powder is deposited onto the existing substrate.  PBF involves a recoater moving and spreading a layer of particles over a base plate, while DED involves powder being blown onto the base plate during the process. In both cases, a laser with a high spatial resolution selectively melts the powder, and the iterative movement of the laser beam over the base plate area followed by a relative movement of the laser with respect to the base plate, results in  layer by layer production of the component.
For capturing the complex phenomena that occur in this process, we selected a set of features that described the surface structure, the packing density of the powder bed, the powder bed morphology and the melt-pool dynamics~\cite{Mindt2016,Parteli2016}. 
%
%
The PBF case study had the input factors: laser power ($W$), scan speed ($mm/s$), powder size ($\mu m$), beam diameter ($mm$), layer thickness ($mm$) and thermal diffusivity ($m^2/s$). A total of approximately 60 samples were analyzed, of which $52\%$ of samples were labeled as defect-free by expert assessment and $48\%$ of the samples presented observable defects. Moreover, we analyzed the performance of the developed approach on seven additional types of metals (for more details about the datasets see~\cite{Du2025}).
The DED case study had the factors: laser power ($W$), scan speed ($mm/s$), powder flow ($rpm$), powder gas ($lpm$), single-scan track length ($mm$), single scan track height ($mm$). Here 36 samples ($40\%$) had identified defects and 45 samples ($50\%$) had no identified defects. Further, there were 9 samples with ``inconclusive'' labels, where it was not possible to recognize either the presence or absence of a defect.

In the interest of reproducibility and replicability, the main code used in this study is publicly available at our GitHub repository: \url{https://github.com/sebabaster/PIGMM-for-AM-/tree/main}.
\section{Discussion and outlook}
%
In this short paper, we present only a few selected result visualizations in Figure~\ref{fig:results}.
We observe the presence of multiple types of data distributions for the different factors in the two processes, including unimodal, bimodal, flattened bell-shaped, and heavy-tailed distributions.
Two examples are illustrated in Figure~\ref{fig:sub1} and  Figure~\ref{fig:sub2}, where the data distribution of the scan speed variable is shown for the L-PBF and DED case studies.
Figure~\ref{fig:sub3} illustrates the diversity of information according to the type of material.
%
%
Figure~\ref{fig:sub4} displays the decision boundary obtained by a Gaussian Mixture Model (GMM), projected onto two relevant variables (power and speed) for the L-PBF case study. This figure visualizes the  capacity of GMMs for predicting defects when the data comes from the L-PBF experiment.
%
The classification performance for the GMMs was different across different alloy types. Suggesting material-specific behavior that influences defect detection. It may also indicate that the models need additional information.
Consequently, several research questions remain open for further studies. We plan to incorporate a multi-resolution hierarchical modeling framework, which can capture various physical phenomena across multiple length scales. Further investigation is still required to analyze the optimal representation of the input information. For instance, we also plan to explore parameter optimization in the frequency domain, as well as other strategies to enhance domain generalization and adaptation~\cite{Basterrech2023, Nguyen2021}.
Another promising direction involves the integration of multi-modal observations, including microtomography imaging and acoustic emission signals. These modalities have shown potential for enhancing defect detection, for example identifying keyhole formation and analyzing crystallization during the AM process.

\subsection*{Acknowledgments}
This work was supported by funding from Horizon Europe, the European Union’s Framework Programme for Research, and Innovation, under Grant Agreement No. 101138859 (DILAPRO). The authors would like to acknowledge the PRIMA Additive by Sodick for providing additional data for the analysis in this paper.
\bibliographystyle{splncs04}
\bibliography{ReferencesMS}
\end{document}